\documentclass[%
manuscript
]{aastex}
\usepackage{graphicx}
\usepackage{epstopdf}
\usepackage{dcolumn}
\usepackage{bm}
\usepackage{hyperref}
\usepackage[all]{hypcap}
\begin{document}


\title{The Cosmic Ray Energy Spectrum Observed with the Surface Detector 
  of the Telescope Array Experiment}

\author{
T.~Abu-Zayyad\altaffilmark{1} 
R.~Aida\altaffilmark{2} 
M.~Allen\altaffilmark{1} 
R.~Anderson\altaffilmark{1} 
R.~Azuma\altaffilmark{3} 
E.~Barcikowski\altaffilmark{1} 
J.W.~Belz\altaffilmark{1}
D.R.~Bergman\altaffilmark{1} 
S.A.~Blake\altaffilmark{1} 
R.~Cady\altaffilmark{1}
% ...
}
\altaffiltext{1}{High Energy Astrophysics Institute and Department of Physics and Astronomy, University of Utah, Salt Lake City, Utah, USA}
\altaffiltext{2}{University of Yamanashi, Interdisciplinary Graduate School of Medicine and Engineering, Kofu, Yamanashi, Japan}
\altaffiltext{3}{Graduate School of Science and Engineering, Tokyo Institute of Technology, Meguro, Tokyo, Japan}
% ...


\begin{abstract} 
The Telescope Array (TA) collaboration has measured the energy
spectrum of ultra-high energy cosmic rays with primary energies above
$1.6\times 10^{18}$~eV.  This measurement is based upon four years of 
observation by the surface detector component of TA.  The
spectrum shows a dip at an energy of $4.6 \times 10^{18}$~eV and a
steepening at $5.4 \times 10^{19}$~eV which is consistent with the
expectation from the GZK cutoff.  
We present the results of a technique, new to the
analysis of ultra-high energy cosmic ray surface detector data, 
that involves generating
a complete simulation of ultra-high energy cosmic rays striking the TA
surface detector.  The procedure starts with shower simulations using
the CORSIKA Monte Carlo program where we have solved the problems
caused by use of the ``thinning'' approximation.  This simulation
method allows us to make an accurate calculation of the acceptance of
the detector for the energies concerned.
\end{abstract}


\maketitle


\section{Introduction}

One of the most powerful tools for studying the origin of ultra-high
energy cosmic rays (UHECRs) is their energy spectrum, which manifests
several features that reveal important information about the cosmic
rays, their sources, and their propagation across cosmological
distances.  One example is the high-energy ($4-6 \times 10^{19}$~eV)
suppression in the spectrum which was predicted by
\citet{greisen} and by \citet{zk},
and is called the GZK cutoff.  These authors predicted a strong
suppression in the spectrum due to the interaction of cosmic rays with
photons of the cosmic microwave background radiation.  They 
pointed out that a spectrum suppression is expected in both cosmic protons
(by photo-pion production) and heavier nuclei (by spallation)
for particles traveling more than 50~Mpc from their sources.
If cosmic rays are protons there should also be a dip in the spectrum,
caused by $e^+e^-$ pair production in the same interactions, at an
energy of about $5 \times 10^{18}$~eV \citep{Berezinsky:1988wi}.  For
heavier nuclei, interactions with the background photon flux do not cause
such a dip. 

The AGASA experiment \citep{agasa:results1,agasa:results2},
comprised of a surface array of 111 scintillation counters, was
the first detector to be large enough to test this theory with sufficient 
statistics. However, they did not
observe the suppression. The first experiment to observe the GZK
cutoff was the High Resolution Fly's Eye (HiRes)
experiment \citep{hiresmono:gzk}, which consisted of fluorescence
detectors located atop two desert mountains in western Utah.  HiRes reported a
cutoff energy of $(5.6 \, \pm 0.5 \, \pm 0.9)
\times 10^{19}$~eV, which is consistent with a suppression of protons.
They also observed the ankle structure: a hardening of the spectrum,
at an energy of $4.5 \, \pm 0.05 \, \pm 0.8 \times 10^{18}$~eV as
expected for cosmic protons.  (For both of these values, the first
uncertainty is statistical and the second systematic.)  HiRes also
published measurements of the shower maximum slant depth ($X_{max}$)
that indicated a predominately light composition above $2 \times
10^{18}$~eV \citep{hires:comp}.

A somewhat different picture is seen by the Pierre Auger Observatory
(PAO), located in Argentina. The PAO consists of a surface detector
(SD) of 1600 water tanks, accompanied by 24 fluorescence telescopes
which are equally apportioned between four sites
located at the SD corners.  The PAO also observes the high-energy
suppression, but at $(2.9 \, \pm 0.2) \times 10^{19}$
eV \citep{auger_gzk,auger:spec}.  They see the ankle also, but
their $X_{max}$ results may indicate that the composition is
heavy at the highest energies \citep{auger:comp}.  
One possible interpretation of the PAO
results is that the high-energy suppression is caused by spallation of
heavy nuclei.  The cause of the ankle would need to be explained by a
separate mechanism.

\begin{figure}[t,b]
  \includegraphics[width=0.9\textwidth]{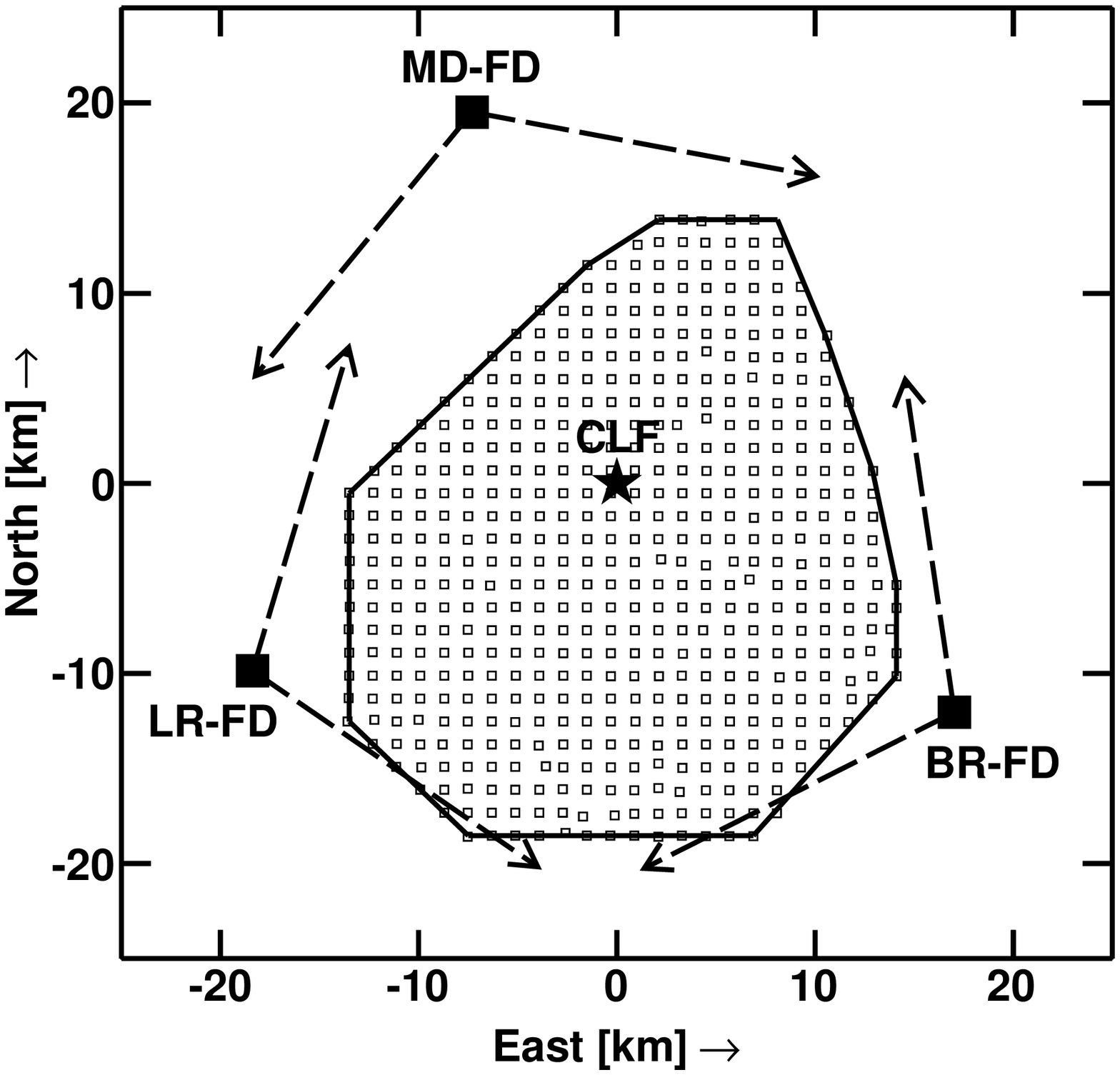}
  \caption{\label{fig:array} Layout of the Telescope Array experiment.
This figure shows the 507 surface detector counters deployed on a 1.2~km grid.
Fluorescence telescopes overlook the surface detector at three sites called 
Middle Drum (MD) on the north, Long Ridge (LR) 
on the southwest, and Black Rock (BR) on the 
southeast. The Central Laser Facility (CLF) is situated in the center of array,
equidistant from all three fluorescence detectors.}
\end{figure}

The Telescope Array (TA) experiment, also in western Utah, is the
largest experiment studying ultra-high energy cosmic rays in the
northern hemisphere.  A layout of the TA experiment is shown in Figure
\ref{fig:array}.
It consists of a surface detector of 507 scintillation
counters \citep{TA:SD}, plus 38 fluorescence
telescopes \citep{TA:FD,TA:MD} located at three sites overlooking
the SD.  24 of the telescopes, deployed at the two southern sites,
were built new for the experiment, and the 14 telescopes at the
northern site are reconditioned HiRes telescopes.  TA combines the
experimental techniques of AGASA and HiRes, in order to understand the
difference between their results.  

This paper reports on a measurement of the cosmic ray spectrum above
$1.6 \times 10^{18}$~eV made by the TA SD over approximately four
years of observation between May 11, 2008 and May 20, 2012.  For this
study, we used an analysis method that while standard for fluorescence
detectors, is being successfully
implemented for the first time for a surface array
studying cosmic rays in the ultra-high ($>1$~EeV) energy regime.  Instead
of restricting our analysis to a domain where we expect 100\%
efficiency, as has been done by previous surface detector experiments at these
energies, the TA SD detector aperture is calculated using extensive
air showers generated in detail by the CORSIKA simulation
package \citep{corsika}, accompanied by a full GEANT simulation of the
detector \citep{geant4}.  Another important aspect of this technique,
new to surface detectors operating in the ultra-high energy regime, 
is the validation of the
simulation by comparisons of key distributions from the data to those
obtained from the Monte Carlo (MC) simulation.  Moreover, our study
overcomes the inability of ``thinned'' simulated showers (e.g. as used
in CORSIKA and AIRES \citep{Sciutto:1999rr}) to reproduce the particle
density and arrival time fluctuations far from the core.  The solution
applied is a novel dethinning technique that replicates a non-thinned
simulation \citep{dethinned} at the lateral core distances where most of the
detector data is collected.

\section{The TA Surface Detector}

Each counter of the TA SD consists of two layers of 1.2 cm thick
plastic scintillator, both 3~m$^2$ in area.  Photons produced by
ionizing particles passing through the counters are collected by
wavelength shifting fibers and read out by photomultiplier tubes, one
for each layer.  A histogram of pulse heights, triggered by a
coincidence between the two layers within an individual SD, is
collected every 10 minutes.  This histogram is dominated by single
muons with a count rate of $\sim 700$~Hz.  Each 10-minute histogram is
used to calibrate the associated scintillator to the pulse height of a minimum
ionizing particle (MIP) with $\sim1\%$ accuracy.  The SD array trigger
requires at least three adjacent counters with pulse heights over 3 MIP
to fire within 8~$\mu$sec.  A 50~MHz flash ADC readout system then saves
the signal traces for all counters in the array with more than 0.3
MIP.  Two fits are used to reconstruct the properties of the cosmic
ray.  First, a fit to the times that counters were struck, using the modified
Linsley shower-shape function \citep{agasa:timefit}, is made to
determine the arrival direction and the core position of the event.
Subsequently, a lateral distribution fit, with the same functional form used by
the AGASA experiment \citep{agasa:results1,agasa:results2}, is
employed to find $S(800)$, the density of shower particles at a lateral
distance of 800~m from the core.  The energy is then estimated by
using a look-up table in $S(800)$ and zenith angle determined from an
exhaustive Monte Carlo simulation.

\section{Aperture Calculation}

In the ultra-high energy regime, computer-time requirements make it
impossible to follow every particle when simulating showers.  An
approximation called thinning is used in programs like CORSIKA and
AIRES to reduce the computational load by only performing a small,
statistically representative sample of the air shower simulation.
Thinned showers can be used for simulation of fluorescence detectors
because the fluorescence light comes mostly from near the shower axis
where the particle density is extremely high, and the fluctuations in
the signal are dominated by the Poisson nature of fluorescence photon
statistics.  But for surface detectors, which operate far from the
shower core, the number of shower particles is low and the thinning
approximation fails to represent the intrinsic density fluctuations
within the shower.  To simulate the TA SD accurately, we have
developed a procedure called ``dethinning,'' where we statistically
regenerate each group of thinned particles from its weighted
representative \citep{dethinned}.

\begin{figure}[t,b]
\includegraphics[width=0.9\textwidth]{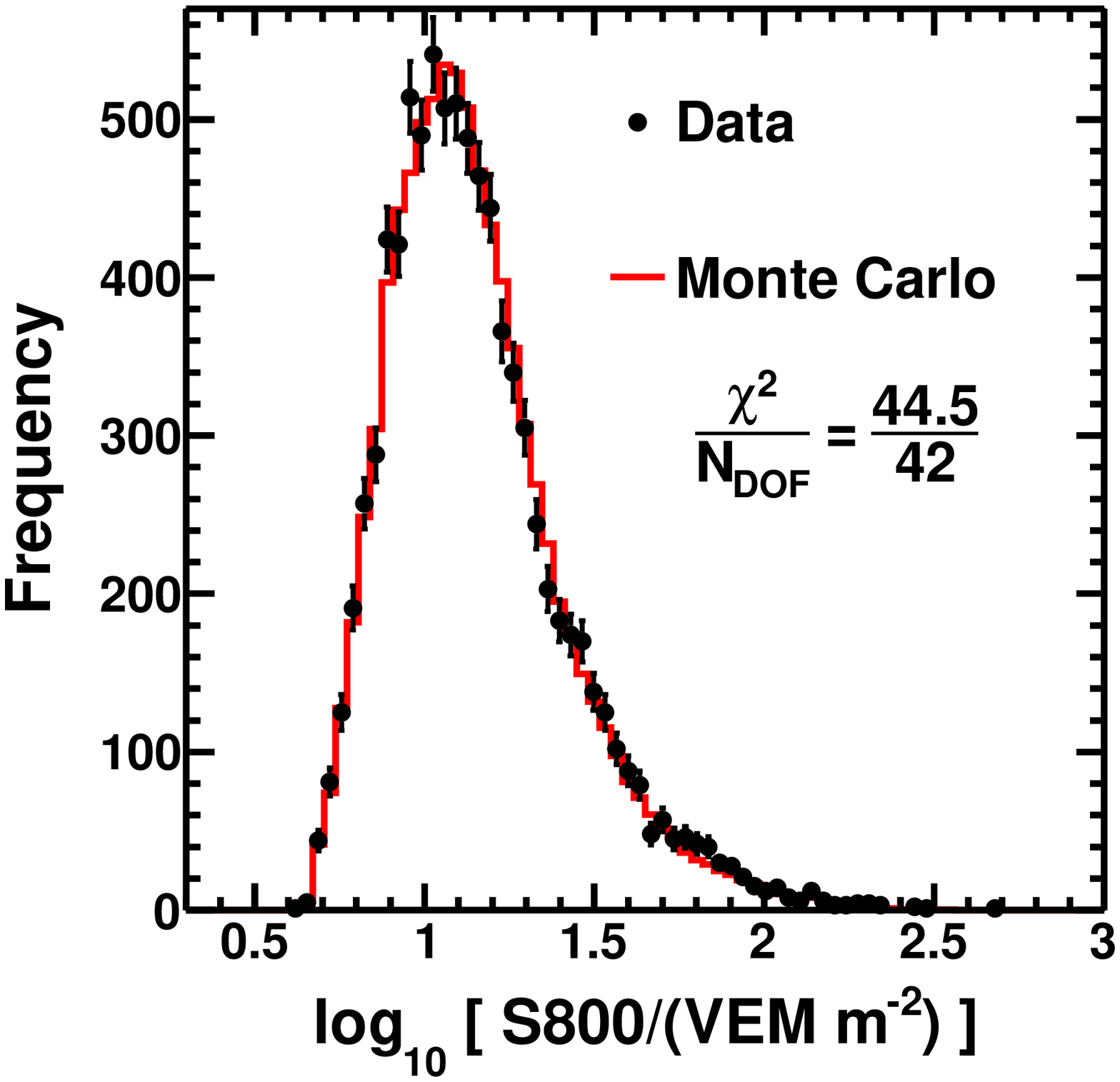}
\caption{\label{fig:dtmc_s800} Data and MC comparison of the event
  $S(800)$ distributions.  The reduced $\chi^{2}$ is 1.06, indicating a good
  agreement. }
\end{figure}

The Monte Carlo simulation of the TA SD has the goal of making an
accurate representation of the data and our detectors.  We start with
a library of showers generated by the CORSIKA program using
QGSJET-II-03 \citep{qgsjet} to model high-energy hadronic interactions,
FLUKA \citep{fluka1,fluka2} to model low-energy hadronic
interactions, and EGS4 \citep{egs4} to model electromagnetic
interactions.    For this library, proton showers
were used exclusively because both the HiRes composition
results \citep{hires:comp} and the preliminary TA composition
result \citep{ta:comp} are consistent with protons generated by
QGSJET-II-03.  A complete representation of calibration and ontime for
each surface counter as a function of time is also included.  Events are 
then chosen from our shower libraries according
to the spectrum previously measured by the HiRes
collaboration \citep{hiresmono:gzk}.Direct
comparisons between data and Monte Carlo show that the result closely
resembles the data \citep{sddtmc}.  Figure~\ref{fig:dtmc_s800}
shows a
comparison of the $S(800)$ of cosmic ray showers.  The excellent agreement
between the data and simulation exemplifies the accuracy of our
simulation and the resulting efficiency calculation of the SD.

\begin{figure}[t,b]
  \includegraphics[width=0.9\textwidth]{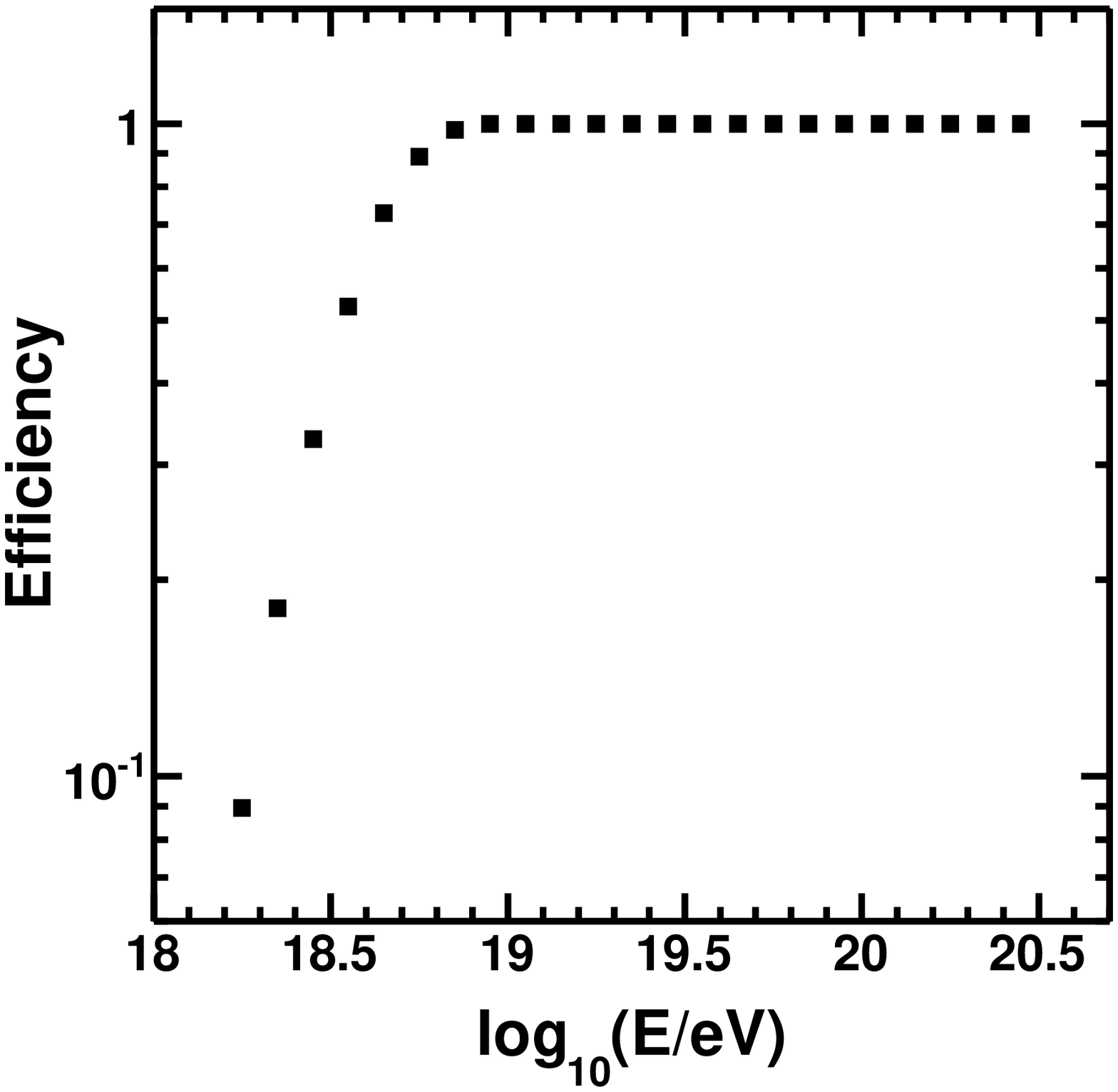}
  \caption{\label{fig:eff_logE} Efficiency as a function of energy.  Both
    trigger and reconstruction effects are included.}
\end{figure}

The selection criteria employed in our analysis are as follows:
\begin{enumerate}
\item Each event must include at least five counters.
\item The reconstructed primary zenith angle must be less than $45^\circ$.
\item The reconstructed event core must be more than 1200~m from edge of the
array.
\item Both the timing and lateral distribution fits must have 
$\chi^2$/degree of freedom value less than 4. 
\item The angular uncertainty estimated by the timing fit must be less than 
$5^\circ$.
\item The fractional uncertainty in $S(800)$ estimated by the lateral distribution
fit must be less than 25\%.
\end{enumerate}
Between May 8, 2008 and May 20, 2012, 13,100 events above $10^{18.2}$ eV were 
collected that satisfy these criteria. 
Figure~\ref{fig:eff_logE}
shows the efficiency of reconstruction calculated from the TA SD Monte
Carlo Program.  The values of aperture and exposure for this data
set, corresponding to the 100\% efficiency region, are
920~$\mathrm{km^{2} \, sr}$ and 3690~$\mathrm{km^{2} \, sr \, yr}$,
respectively.  For energies above $10^{18.2}$~eV (where the efficiency
is $\sim 10\%$ of its plateau value) we can accurately simulate all air
showers, both well- and poorly- reconstructed.  The resolution of the
TA SD energy determination is better than 20\% above $10^{19}$~eV.

\begin{figure}[t,b]
  \includegraphics[width=0.9\textwidth]{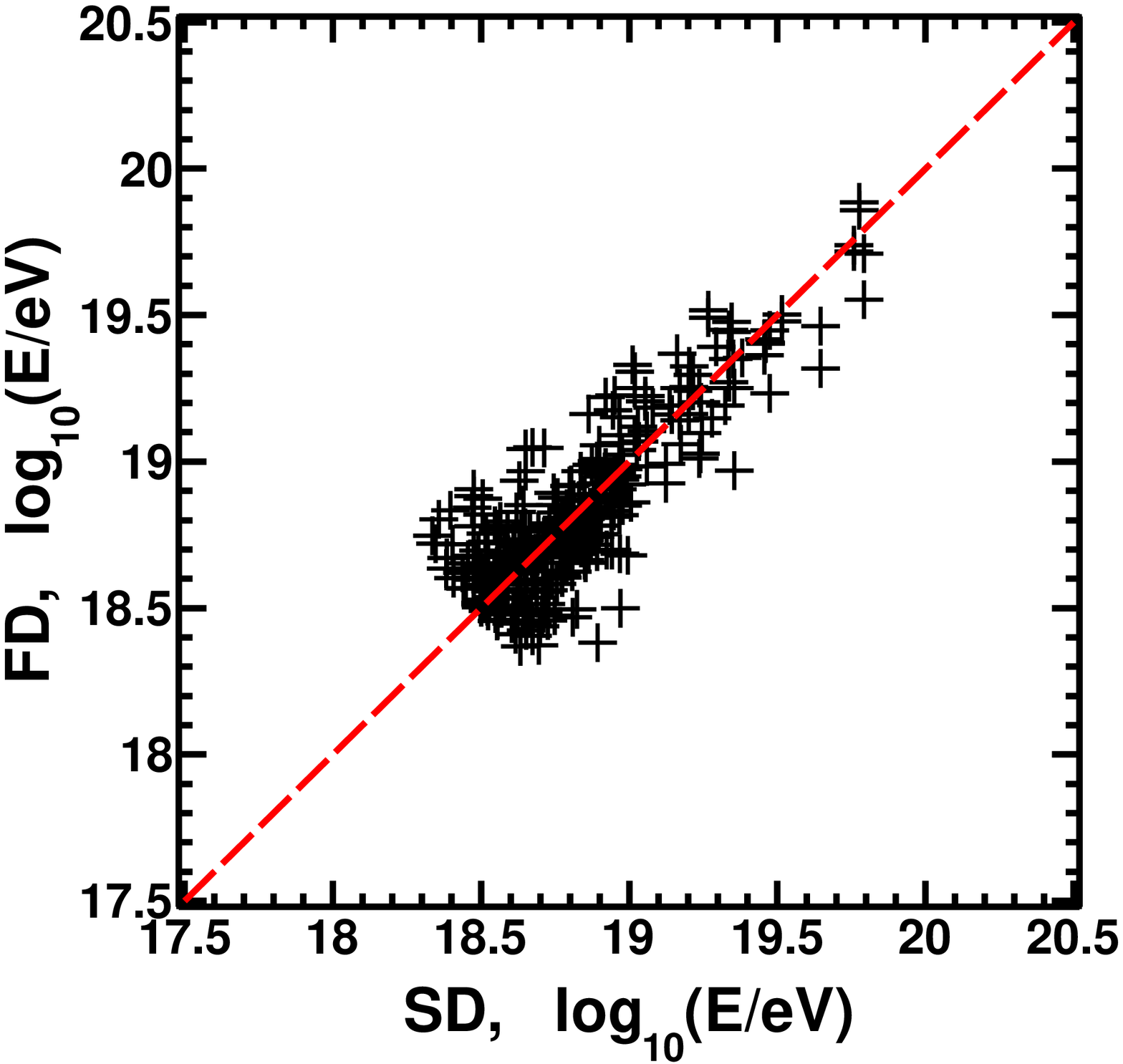}
  \caption{\label{fig:en_scatter} Energy comparison between the
    TA SD and FD \emph{after} the 27\% normalization has been applied to the SD. }
\end{figure}

The uncertainty in energy scale of the Monte Carlo simulation of an SD
is large, and possible biases associated with the modeling of hadronic
interactions (e.g., from extrapolations of cross sections measured at much
lower energies) are difficult to determine.  However, the energy scale
uncertainty is experimentally well-controlled for a fluorescence
detector (FD) since the energy measurement is calorimetric.  We
therefore correct our energy scale to the TA FD using events seen in
common between the FD and SD.  The observed differences between the FD
and SD events are well described by a simple proportionality
relationship, where the SD energy scale is 27\% higher than the FD.
Figure~\ref{fig:en_scatter}
shows a scatter plot of FD vs SD energies, where the latter have
been rescaled.  Events from all three FD stations were included in
this plot.  The two southern FD stations were calibrated using
independent techniques from the northern station, which consists of
reconditioned HiRes fluorescence telescopes.  The resulting energy
scales are consistent for all TA fluorescence detectors.

\section{Spectrum}

\begin{figure}[t]
  \includegraphics[width=0.9\textwidth]{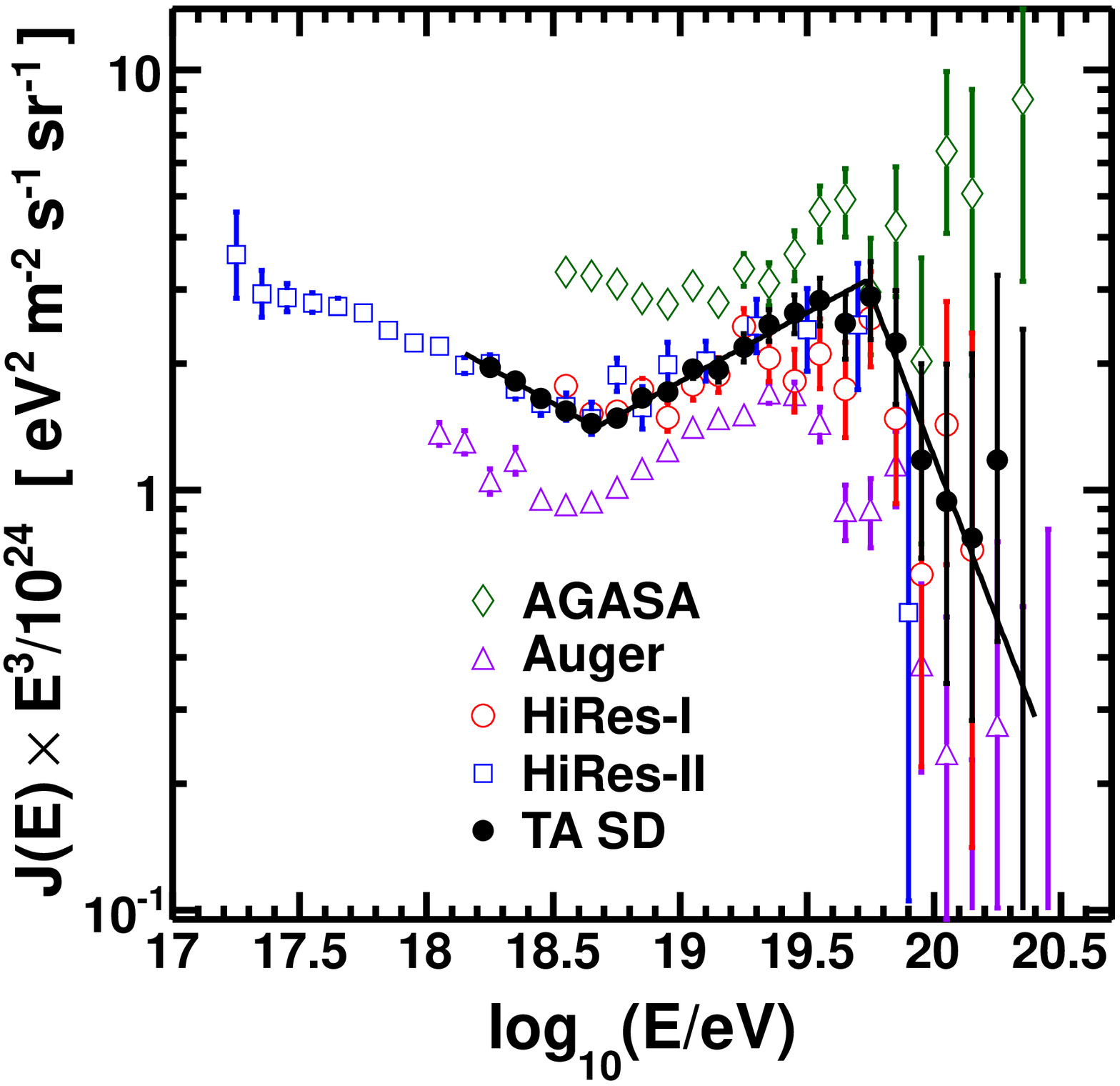}
  \caption{\label{fig:e3j} Cosmic ray flux multiplied by E$^3$. Solid
    line shows the fit of the TA SD data to a broken power law. }
\end{figure}

Figure~\ref{fig:e3j} 
shows the spectrum measured by the TA SD, where the differential flux,
$J(E) = \mathrm{d}^{4}N(E) \, / \, \mathrm{d}E \, \mathrm{d}A \,
\mathrm{d}\Omega \, \mathrm{d}t$ is multiplied by $E^3$, and plotted
against $\mathrm{log_{10}}E$. The ankle structure and the suppression
at the highest energies are clearly visible.  A fit to a broken power
law (BPL) determines the energies of these features. 
The fit finds the ankle at an energy of 
$(4.6\pm 0.3) \times 10^{18}$~eV and the suppression at $(5.4 \pm 0.6)
\times 10^{19}$~eV. The power exponents for the three regions (below
the ankle, between the breaks, and above the suppression) are $-3.34
\pm 0.04, -2.67 \pm 0.03$, and $-4.6 \pm 0.6$ respectively.  Also
shown in Figure~\ref{fig:e3j} are the spectra reported by
AGASA \citep{agasa:results2}, HiRes (monocular
mode) \citep{hiresmono:gzk}, and PAO (combined hybrid and
SD) \citep{auger:spec}. The HiRes and TA SD spectra agree very well,
both in the energy region above $10^{18.85}$ eV where the TA SD is
100\% efficient, and also at lower energies where TA employs a substantial
efficiency correction.

A linear extrapolation of the power law below the suppression predicts
58.6 events above the break; whereas TA observed only 21 events.
This difference corresponds to a Poisson probability of $1.44 \times
10^{-8}$, or 5.5 standard deviations significance.  A related observable,
$E_{1/2}$, is the
energy at which the integral spectrum falls to 1/2 of its expected
value in the absence of the GZK cutoff.  Under a wide range of
assumptions about the spectrum of extragalactic sources, $E_{1/2}$ is
predicted to be $10^{19.72}$ eV for protons \citep{Berezinsky:2002nc}.
HiRes reported $\log_{10} E=19.73 \pm 0.07$ \citep{hiresmono:gzk}.  We
measure $\log_{10} E=19.72 \pm 0.05$.

This 5.5 standard deviation observation provides independent
confirmation of the GZK cutoff observed by HiRes \citep{hiresmono:gzk}.
Furthermore, the energy of the cutoff is consistent with the
interpretation that the composition is protonic.

\citet{sddtmc} includes a description of systematic
uncertainties in the SD spectrum measurement.  The largest source of
systematic uncertainty in the spectrum is that of the energy scale.
Since the SD energy scale is fixed to that of the TA fluorescence
detectors, we take the systematic uncertainty in the SD energy to be
22\% \citep{ta:enscale_sys}, the same as the FD.  This propagates into
a 37\% uncertainty in the flux.  We estimate the systematic
uncertainty in the aperture calculation by removing the event
selection criteria, one by one, and measuring the ratio of the number
of events in the data and in the Monte Carlo simulation.  This ratio
does not change by more than 3\% in any energy bin above $10^{18.2}$
eV, so we assign this value to be the systematic uncertainty in the
aperture.

\section{Conclusions}

We have measured the spectrum of cosmic rays in the energy range
$10^{18.2} - 10^{20.3}$~eV using the surface detector of the Telescope
Array experiment.  In the analysis, we have introduced a technique,
new to the ultra-high energy regime for surface detectors, 
of calculating the surface
detector aperture using Monte Carlo simulation, which allows us to
measure the spectrum even when the SD efficiency is less than 100\%.
This technique includes a dethinning process that enables the
simulation of air showers with excellent detail.  We found that the
energy scale of the SD determined from simulations can be reconciled
with the calorimetric scale of fluorescence detectors by a simple
renormalization of 27\%.

Two features are seen in the spectrum, the ankle and the high-energy
suppression.  Fitting the spectrum to a broken power law shows a
definite break at an energy of $(5.4\pm 0.6)\times 10^{19}$~eV, which
is consistent with the GZK cutoff energy expected for protons.  An
extended spectrum beyond the GZK energy is ruled out with a
statistical significance of 5.5 standard deviations.  Our result is in
excellent agreement with that of the HiRes experiment where
fluorescence detectors were used.  This result demonstrates,
contrary to the AGASA claim, that there
is no difference between measurements of the cosmic ray spectrum using
a fluorescence detector and a surface scintillation array once the
energy scales are normalized.  

In summary, by combining the two techniques of surface detectors and
fluorescence detectors (used by the AGASA and HiRes experiments), we
have now obtained a consistent energy spectrum for ultra-high energy
cosmic rays from both techniques.  The spectrum obtained by our
experiment demonstrates spectral features, a dip and a cutoff,
consistent with the interaction of extra-galactic protons with the
cosmic microwave background (GZK process).  Finally, if we account for a 20\%
systematic difference in energy scale, our measurement is in good agreement 
with the spectrum reported by PAO with one exception: the GZK break is 
reported at $(2.9\pm 0.2)\times 10^{19}$~eV
by PAO \citep{auger:spec}; even with a 20\% energy scale correction the 
difference between the TA and PAO measurements is three standard deviations.

\acknowledgments

The Telescope Array experiment is supported 
by the Japan Society for the Promotion of Science through
Grants-in-Aids for Scientific Research on Specially Promoted Research (21000002) 
``Extreme Phenomena in the Universe Explored by Highest Energy Cosmic Rays''
and for Scientific Research (S) (19104006),  
and the Inter-University Research Program of the Institute for Cosmic Ray 
Research;
by the U.S. National Science Foundation awards PHY-0307098, 
PHY-0601915, PHY-0703893, PHY-0758342, and PHY-0848320 (Utah) and 
PHY-0649681 (Rutgers); 
by the National Research Foundation of Korea 
(2006-0050031, 2007-0056005, 2007-0093860, 2010-0011378, 2010-0028071, R32-10130, 2011-0002617);
by the Russian Academy of Sciences, RFBR
grants 10-02-01406a and 11-02-01528a (INR),
IISN project No. 4.4509.10 and 
Belgian Science Policy under IUAP VI/11 (ULB).
The foundations of Dr. Ezekiel R. and Edna Wattis Dumke,
Willard L. Eccles and the George S. and Dolores Dore Eccles
all helped with generous donations. 
The State of Utah supported the project through its Economic Development
Board, and the University of Utah through the 
Office of the Vice President for Research. 
The experimental site became available through the cooperation of the 
Utah School and Institutional Trust Lands Administration (SITLA), 
U.S.~Bureau of Land Management and the U.S.~Air Force. 
We also wish to thank the people and the officials of Millard County,
Utah, for their steadfast and warm support. 
We gratefully acknowledge the contributions from the technical staffs of our
home institutions.  An allocation of computer time from the Center for High Performance Computing at the University of Utah is gratefully acknowledged. 


\end{document}